\pdfoutput=1
\documentclass[fleqn,letters,usenatbib,nofootinbib]{mnras}
\usepackage{graphicx}
\usepackage{amsmath,amssymb}
\usepackage{lastpage}
\usepackage[all]{hypcap}

\usepackage[T1]{fontenc}

\usepackage{float}
\usepackage{subfigure}
\usepackage{afterpage}
\usepackage[usenames]{color}
\usepackage{yfonts}
\usepackage{mathrsfs}
\usepackage{upgreek}
\usepackage{hyperref}

\newcommand{\be}{\begin{equation}}
\newcommand{\ee}{\end{equation}}
\newcommand{\bea}{\begin{eqnarray}}
\newcommand{\eea}{\end{eqnarray}}

\usepackage[stable]{footmisc}
\usepackage{color}

\def\tt{\Uptheta}

\def\dkmu2{\delta K_{\mu \nu}\delta K^{\mu \nu}}
\def\pmu2{  \phi_{\mu \nu}\phi^{\mu \nu}}

\pubyear{2018}

\begin{document}
\label{FirstPage}
\pagerange{\pageref{FirstPage}--\pageref{LastPage}}

\title{The Fifth Force in the Local Cosmic Web}

\author[H.~Desmond, P.~G.~Ferreira, G.~Lavaux, J.~Jasche]{
Harry~Desmond$^{1}$\thanks{E-mail: harry.desmond@physics.ox.ac.uk},
Pedro~G.~Ferreira$^{1}$,
Guilhem~Lavaux$^{2,3}$,
and \newauthor{}
Jens~Jasche$^{4,5}$
\\
$^{1}$Astrophysics, University of Oxford, Denys Wilkinson Building, Keble Road, Oxford OX1 3RH, UK\\
$^{2}$Sorbonne Universit{\'e}, CNRS, UMR 7095, Institut d'Astrophysique de Paris, 98 bis bd Arago, 75014 Paris, France\\
$^{3}$Sorbonne Universit{\'e}s, Institut Lagrange de Paris (ILP), 98 bis bd Arago, 75014 Paris, France\\
$^{4}$The Oskar Klein Centre, Department of Physics, Stockholm University, Albanova University Center, SE 106 91 Stockholm, Sweden\\
$^{5}$Excellence Cluster Universe, Technische Universit{\"a}t M{\"u}nchen, Boltzmannstrasse 2, D-85748 Garching, Germany\\
}
    
\maketitle

\begin{abstract}
Extensions of the standard models of particle physics and cosmology often lead to long-range fifth forces with properties dependent on gravitational environment. Fifth forces on astrophysical scales are best studied in the cosmic web where perturbation theory breaks down. We present constraints on chameleon- and symmetron-screened fifth forces with Yukawa coupling and megaparsec range -- as well as unscreened fifth forces with differential coupling to galactic mass components -- by searching for the displacements they predict between galaxies' stars and gas. Taking data from the \textit{Alfalfa} H\textsc{i} survey, identifying galaxies' gravitational environments with the maps of~\citet{Desmond:2017ctk} and forward-modelling with a Bayesian likelihood framework, we set upper bounds on fifth-force strength relative to Newtonian gravity from $\Delta G/G_N < \text{few} \: \times 10^{-4}$ for range $\lambda_C = 50$ Mpc, to $\Delta G/G_N \lesssim 0.1$ for $\lambda_C = 500$ kpc. In $f(R)$ gravity this requires $f_{R0} < \text{few} \: \times \: 10^{-8}$. The analogous bounds without screening are $\Delta G/G_N < \text{few} \: \times 10^{-4}$ and $\Delta G/G_N < \text{few} \times 10^{-3}$. These are the tightest and among the only fifth-force constraints on galaxy scales. We show how our results may be strengthened with future survey data and identify the key features of an observational programme for furthering fifth-force tests beyond the Solar System.
\end{abstract}


\section{Introduction}
\label{sec:intro}

Despite fundamental open questions, almost all attempts at extending the standard models of particle physics and cosmology have proven unsatisfactory. Nevertheless, a generic feature of such extensions is the introduction of extra degrees of freedom. These arise by replacing dimension-full parameters with dynamical fields [e.g. lepton masses~\citep{Weinberg:1967tq}, dynamical dark energy~\citep{Ratra:1987rm} or the gravitational constant~\citep{Brans:1961sx,Wetterich:1987fm}], and embody higher derivatives and extra dimensions. As any generalisation of the Einstein-Hilbert action must evolve new fields~\citep{Clifton:2011jh}, practically all attempts to extend the standard model add scalar, vector or tensor fields that influence the dynamics of the Universe and its contents.

Extra fields couple naturally to the Ricci scalar $R$ in the gravitational action. For example, a scalar $\phi$ may generate a non-minimal coupling of the form $\phi^2 R$, which complicates dynamics: not only will it source energy and momentum (along with all other constituents of the Universe) but it will also modify the gravitational force. Taking the simplest case of standard kinetic energy and potential $V(\phi)$, the Newtonian potential $\Phi$ of a point mass $M$ is modified to
\begin{eqnarray}
\Phi_\text{tot}=\frac{G M}{r}\left(1+\frac{\Delta G}{G}e^{-mr}\right)  \label{ff}
\end{eqnarray}
where $G$ is the bare (Newtonian) gravitational constant, $m\sim d^2V/d\phi^2$ and $\Delta G/G$ depends on the magnitude of the non-minimal coupling and the background field value relative to the Planck mass $M_\text{pl}$. $m$ sets the range of the fifth force and $\Delta G$ its strength. The General Relativistic (GR) result is recovered for $\Delta G\rightarrow0$, and also for $m\rightarrow\infty$ so that the fifth force is confined to a narrow radius around the source. The scalar Higgs field for example generates a very short-range fifth force~\citep{Herranen:2015ima}.
\begin{figure}
\vspace{-20mm}
  \centering
  \includegraphics[width=0.5\textwidth]{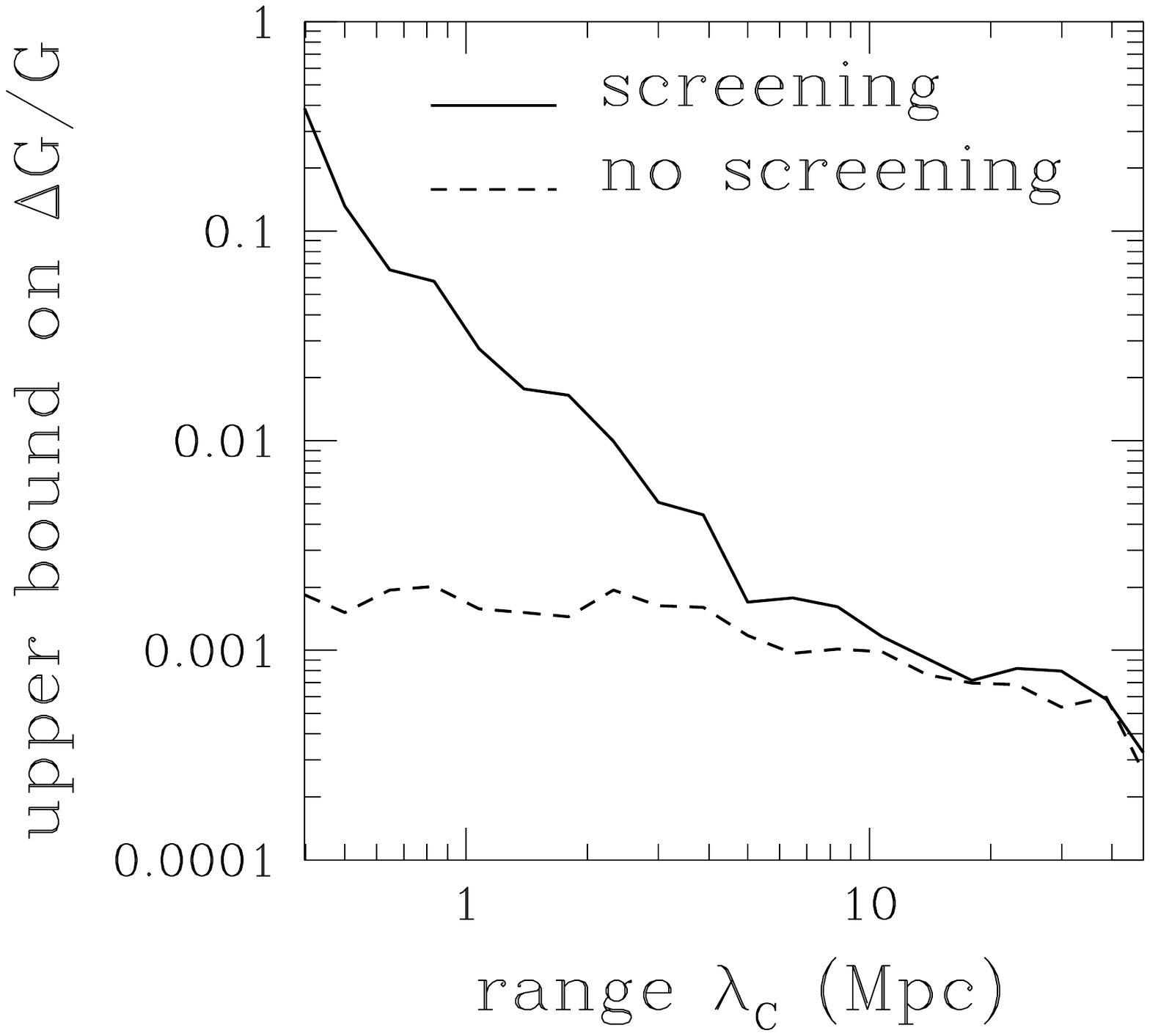}
  \vspace{-40mm}
  \caption{$1\sigma$ bound on fifth-force strength $\Delta G/G$ and range $\lambda_C$ obtained from offsets between the stellar and gas mass centroids of \textit{Alfalfa} galaxies, both with and without screening.}
  \label{fig:constraint}
   \vspace{-5mm}
\end{figure}

There are extremely stringent constraints on fifth forces over a wide range of scales (see~\citealt{Adelberger:2003zx} for a review); on astrophysical scales the tightest constraints for low $m$ come from Shapiro time delay measurements from the Cassini satellite~\citep{Bertotti:2003rm}, which require $\Delta G/G \lesssim 10^{-5}$. Although this is sufficiently strong to make a universally-coupled fifth force cosmologically insignificant, a number of theories (for example generalised scalar-tensor theories and massive gravity) evade Solar System bounds by means of a \textit{screening mechanism} whereby the fifth-force strength or range becomes a function of environment. {\it Chameleon} screening~\citep{Khoury:2003rn} arises when the effective mass, $m_{\rm eff}$, becomes dependent on local density (and thus on $\nabla^2\Phi$, where $\Phi$ is the Newtonian potential): in denser regions, $m_\text{eff}$ become large and the fifth force has short range, while in empty regions (or on cosmological scales) $m_{\rm eff}\rightarrow0$ and the fifth force effectively emerges. By virtue of the `thin-shell effect', and corroborated in simulations~\citep{Zhao_1, Zhao_2, Cabre}, an object's degree of screening is set by $\Phi = \Phi_\text{in} + \Phi_\text{ex}$, where $\Phi_\text{in}$ is the potential at the object's surface due to its own mass and $\Phi_\text{ex}$ is the contribution from surrounding mass. The object is unscreened if $|\Phi|$ is less than a critical value $|\Phi_\text{c}|$. Conversely, in the {\it Vainshtein}~\citep{Vainshtein:1972sx} and {\it symmetron}~\citep{Hinterbichler:2010es} mechanisms the fifth-force strength depends on environment: near massive bodies $\Delta G/G\rightarrow 0$, while away from them $\Delta G/G\neq 0$.

In the presence of screening, the laboratory, the Solar System and clusters will generally probe the screened regime and hence be expected to yield the GR result. However, this is not the case for a range of galaxy environments in the cosmic web, which probe very low density regions and should therefore manifest a fifth force. In this Letter we use a map of screening proxies to identify these environments and hence forward-model a key signal of chameleon and symmetron screening: a displacement between galaxies' stellar and gas mass centroids. Comparing to optical and H\textsc{i} data, we set $1\sigma$ limits from $\Delta G/G < \text{few} \: \times 10^{-4}$ at range $1/m_\text{eff} \simeq \lambda_C = 50$ Mpc to $\sim 0.1$ for $\lambda_C = 500$ kpc. In $f(R)$ gravity, where $\Delta G/G = 1/3$, this corresponds to $f_{R0} \lesssim \: \text{few} \: \times \: 10^{-8}$.

\section{Methods and observables}
\label{sec:method}

The detailed procedure for charting the gravitational environments of the local Universe is given in~\citet{Desmond:2017ctk} (building on earlier work in~\citealt{Cabre}); we provide a summary here. Our map encompasses a region out to approximately $200 \: h^{-1}$ Mpc and is based on the 2M++ galaxy catalogue~\citep{2M++}, a synthesis of 2MASS, 6dF and SDSS data. We connect the $K$-band luminosity function with the halo mass function from a high resolution $\Lambda$CDM N-body simulation (\textsc{darksky-400};~\citealt{darksky}) by using abundance matching (AM) to associate a dark matter halo to each galaxy, according to the specific prescription of~\citealt{Lehmann}. (We validate this model in the $K$-band using a counts-in-cells clustering statistic in~\citet{Desmond:2017ctk}.) The magnitude limit of the 2M++ survey (12.5 in $K$) means that it misses faint galaxies and their associated halos. To correct for this, we use the abundance-matched simulation to estimate the distribution and density of halos hosting galaxies above the magnitude limit, and fill these in through their probabilistic correlation with observables. Finally, we account for the matter not associated with resolved halos by means of a Bayesian reconstruction of the density field with resolution $2.65 \: h^{-1}$ Mpc using the BORG algorithm~\citep{BORG_1, BORG_2, BORG_3, BORG_PM}, which propagates information from the number densities and peculiar velocities of 2M++ galaxies assuming concordance cosmology and a bias model. We call this the ``smooth density field''. As each step in this chain is probabilistic, we generate many Monte Carlo realisations of the fields to sample the statistical uncertainties in the inputs.

We focus here on a particular fifth-force signal: the displacement between galaxies' optical (tracing stellar mass) and H\textsc{i} (tracing cold gas mass) centroids. Such a displacement may come about either from a difference in the coupling of the fifth force to stars and gas, or, more likely, from chameleon or symmetron screening~\citep{Jain_Vanderplas, Brax_2}. In the latter, gas and dark matter in unscreened galaxies feel a fifth force due to neighbouring unscreened mass, leading to an effective increase in Newton's constant $\Delta G = 2\beta^2 G$ for coupling coefficient $\beta$ if the scalar field is light. Stars on the other hand self-screen and feel only $G$. The result of this effective equivalence principle violation~\citep{Hui} is an offset between the stellar and gas mass in the direction of the external fifth-force $\vec{a}_5$.

We search for such a displacement, and its correlation with $\vec{a}_5$, using the complete catalogue of \textit{Alfalfa}~\citep{Giovanelli, Kent, Haynes11}, a blind H\textsc{i} survey out to $z \simeq 0.06$ conducted with the Arecibo observatory. Optical counterparts (OCs) for the majority of detections were derived from cross-correlation with optical surveys and included in the catalogue. The uncertainty in the H\textsc{i} centroid position is best estimated directly from its displacement from the OC: we create 50 logarithmically uniform bins in the signal to noise ratio of the detection (SNR) between the minimum and maximum values 4.6 and 1000 respectively, calculate in each bin the standard deviation of the RA and DEC components of the H\textsc{i}-optical offset, and set the corresponding components of the H\textsc{i} centroid uncertainties to be twice these to ensure our constraints are conservative. This gives the uncertainty a median and standard deviation across the sample of $36\arcsec$ and $8\arcsec$ respectively. (We briefly mention the results of a less conservative choice below, and note that similar results are obtained by fitting for the uncertainty as a zeroth, first or second order polynomial in SNR.) We cut the catalogue at 100 Mpc where the fixed angular uncertainty leads to an unacceptably large spatial uncertainty, yielding a sample of size $12,177$. We then cut a further 1,355 galaxies with poor SNR (\textit{Alfalfa} quality flag 2 or 9) and 262 galaxies where the optical and H\textsc{i} images are likely misidentified ($>2\arcmin$ H\textsc{i}-OC offset), which corresponds roughly to a $3\sigma$ outlier clip. We have checked that our analysis is not especially sensitive to this: even cutting at $1\arcmin$ (a $<$2$\sigma$ clip), removing $4.6\%$ of our sample, does not appreciably alter our results. Our final sample has size $N_\text{Alf} = 10,822$. We supplement the \textit{Alfalfa} information for $22\%$ of our galaxies with structural galaxy properties from the \textit{Nasa Sloan Atlas} (NSA; stellar mass $M_*$, half-light radius $R_\text{eff}$, apparent axis ratio $b/a$, aperture velocity dispersion $\sigma_\text{d}$ and S\'{e}rsic index $n$), which will improve the precision of the predicted H\textsc{i}-OC offset as calculated below.

To constrain the fifth-force strength $\Delta G$ and range $\lambda_C$ we proceed as follows. First, assuming a Compton wavelength for the scalar field in the range $0.4 < \lambda_C/\text{Mpc} < 50$ we set the screening threshold
\begin{equation} \label{eq:f(R)}
|\Phi_\text{c}|/c^2 = \frac{3}{2} \times 10^{-4} \left(\frac{\lambda_C}{32 \: \text{Mpc}}\right)^2.
\end{equation}
This is exact for the case of Hu-Sawicki $f(R)$~\citep{Hu:2007nk} (where $|\Phi_\text{c}|$ is $1.5$ times the background scalar field value $\phi_0 = f_{R0}$) and also applicable more generally with $\lambda_C$ interpreted in terms of the self-screening parameter $\phi_0/(2\beta M_\text{pl})$. We use our gravitational maps to determine which halos, and portions of the smooth density field, are unscreened given these parameters by calculating $\Phi_\text{ex}$ as a sum over all mass within $\lambda_C$ of the test point. We take $\Phi_\text{in} = -\sigma_\text{d}^2$ for galaxies with NSA information and $\Phi_\text{in} = -V_\text{max}^2$ for those without, where $V_\text{max}$ is the maximum rotational velocity estimated by correcting the full-width half-max of the radio detection for turbulence and projection effects~\citep{Tully}. These contributions to the total potential derive primarily from the test galaxy's dark matter. Note that in the case of cluster galaxies, the potential of the cluster itself is part of the external contribution. We calculate $\vec{a}_5$ by summing the contributions of all unscreened mass within $\lambda_C$. We then calculate the equilibrium H\textsc{i}-OC offset $\vec{r}_*$ predicted for a given galaxy:
\begin{equation} \label{eq:rstar}
\frac{M(<r_*)}{r_*^2} \: \hat{\vec{r}}_* = \frac{\Delta G}{G^2} \: \vec{a}_5
\end{equation}
if it is unscreened and $0$ otherwise, where $M(<r_*)$ is the dark matter plus gas mass between the H\textsc{i} and optical centroids. This follows from the requirement that the extra force on the stellar disk due to its offset from the halo centre compensate for its not feeling the fifth force, so that the stars, gas and dark matter continue to move together~\citep{Jain_Vanderplas}. We calculate $M(<r_*)$ by assuming a constant density $\rho_0$ within $r_*$ (justified post-hoc: $r_*$ for the fifth-force models we are sensitive to is $10^{-2}-10^{-1} \: \text{kpc}$, much less than the halo scale radius $r_s$), and estimate it separately for each galaxy using the empirical relation between central baryonic and dynamical surface mass densities~\citep{Lelli_1, Lelli_2, Milgrom}. This yields
\begin{equation} \label{eq:sig}
\vec{r}_* = \frac{3}{4\pi}\frac{1}{\rho_0}\frac{\Delta G}{G^2} \: \vec{a}_5.
\end{equation}
As $\vec{r}_*$ spans a very small angle on the plane of the sky we compare separately its orthogonal RA ($r_{*,\alpha}$) and DEC ($r_{*,\delta}$) components with those of the measured displacement for each galaxy.
\begin{figure}
\vspace{-20mm}
  \centering
  \includegraphics[width=0.5\textwidth]{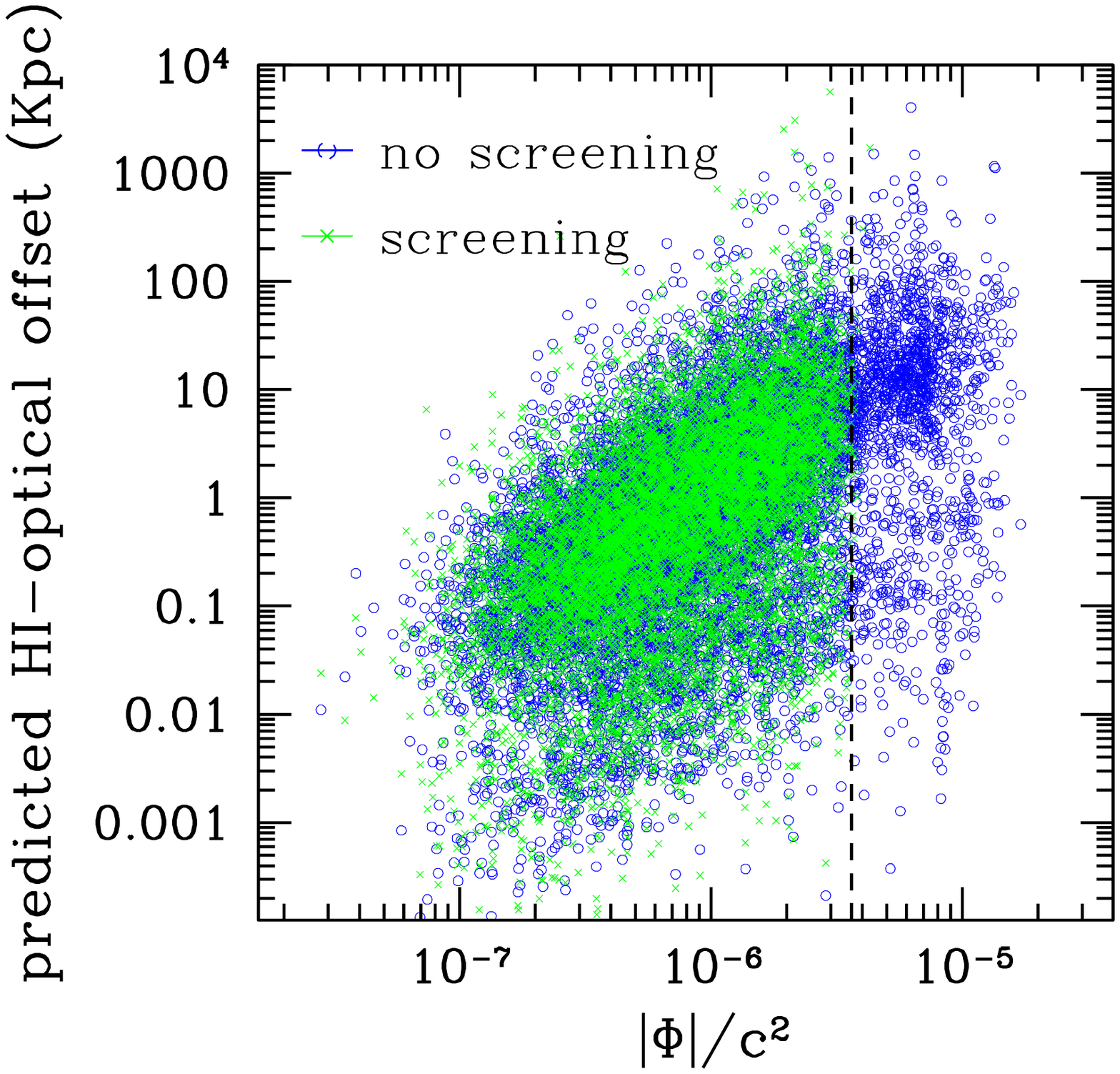}
  \vspace{-40mm}
  \caption{Offsets $r_*$ between optical and H\textsc{i} centroids predicted for \textit{Alfalfa} galaxies within 100 Mpc by a model with $\lambda_C = 5$ Mpc and $\Delta G/G = 1$, as a function of total Newtonian potential $\Phi$. The green points are for the full model with screening; the blue points show the case where screening is switched off. The bars in the legend show the average size of the uncertainties in $\Phi$, defined as the minimal widths enclosing 68\% of the Monte Carlo realisations of the model. The y-uncertainties are too small to be visible on this plot, and are subdominant to the measurement uncertainties. The vertical dashed line shows the threshold $|\Phi_\text{c}|$ above which galaxies in the model with screening are screened.}
  \label{fig:scatter}
\end{figure}

We feed these calculations into a Bayesian likelihood formalism. First, we generate $N_\text{MC}=1000$ Monte Carlo realisations of the predicted signal $\vec{r}_*$ for each \textit{Alfalfa} galaxy, sampling independently for each one the galaxy--halo connection (from 200 independent AM realisations), the distribution of mass in the smooth density field (from 10 particle-mesh BORG realisations), the contribution to $\Phi_\text{ex}$ and $\vec{a}_5$ from halos too faint to be recorded in 2M++ (calibrated with the \textsc{darksky-400} N-body box), and the Gaussian observational uncertainties on the structural galaxy properties used to derive $M(<r_*)$ and $\Phi_\text{in}$. The full probability distributions that we marginalise over are given in table 1 of~\citet{Desmond_PRD}. We estimate the probability that a given galaxy is unscreened as $f \equiv N(|\Phi_\text{ex}| + |\Phi_\text{in}| < |\Phi_\text{c}|)/N_\text{MC}$. The likelihood function then has separate screened ($r_*=0$) and unscreened (Eq.~\ref{eq:sig}) components, with relative weights $1-f$ and $f$ respectively. We model the unscreened component using a normalised histogram of the distributions of $r_{*,\alpha}$ and $r_{*,\delta}$ over all $N_\text{MC}$ realisations, obviating the need for assumptions on the form of the likelihood function such as Gaussianity. We convolve this likelihood with the Gaussian H\textsc{i} measurement uncertainty for each galaxy, $\theta_i$, and treat galaxies as uncorrelated and RA and DEC components as independent. This gives the total likelihood of the \textit{Alfalfa} data under the fifth-force model specified by $\{\lambda_C, \Delta G\}$. Finally, we take 20 logarithmically uniformly spaced values of $\lambda_C$ between 400 kpc and 50 Mpc and constrain $\lambda_C$ and $\Delta G/G$ by MCMC.

Our study greatly extends previous work testing chameleon screening by means of this signal~\citep{Vikram}, in which $M(<r_*)$ and $\vec{a}_5$ were not modelled. We describe our method exhaustively in~\citet{Desmond_PRD}.

\section{Results}
\label{sec:results}

In Fig.~\ref{fig:constraint} we show our $1\sigma$ constraint in the $\lambda_C-\Delta G/G$ plane, with and without screening. The dependence of the $\Delta G/G$ limit on $\lambda_C$ may be understood as a combination of two effects. First, when $\lambda_C$ is smaller less mass contributes to $\vec{a}_5$, leading to a smaller predicted signal at fixed $\Delta G/G$ (Eq.~\ref{eq:sig}). This allows $\Delta G/G$, which simply scales the predicted $\vec{r}_*$, to be larger while keeping the prediction consistent with the observations. Second, a smaller $\lambda_C$ corresponds to a smaller $|\Phi_\text{c}|$ (Eq.~\ref{eq:f(R)}), making both the test galaxy itself and the surrounding mass less likely to be unscreened, and hence to contribute to $\vec{a}_5$. In the case without screening, $\vec{a}_5$ is calculated from \emph{all} mass within $\lambda_C$ (rather than only unscreened mass), and each test galaxy is considered fully unscreened ($f=1$). Removing screening strengthens the $\Delta G/G$ constraints at low $\lambda_C$ but does not change them significantly for $\lambda_C \gtrsim 10$ Mpc, because at higher $\lambda_C$ most masses are unscreened anyway. Instead, the factor limiting the constraint is the volume around a galaxy within which matter contributes to $\vec{a}_5$, which is set by $\lambda_C$ and is the same between the screening and no-screening runs. Similar results are obtained from resamples of the \textit{Alfalfa} data with repeats (bootstraps) and from parts of the full dataset (jackknifes).

In Fig.~\ref{fig:scatter} we show the correlation with $\Phi$ of the signal $r_*$ predicted for the \textit{Alfalfa} galaxies by a fiducial model with $\lambda_C = 5$ Mpc, $\Delta G/G = 1$. Green points are for the case with screening included (so that $r_* \rightarrow 0$ for $|\Phi| > |\Phi_c|$) and blue for the case without. For this relatively high value of $\Delta G/G$ the predicted signal is typically $\mathcal{O}$(kpc). The trend with $\Phi$ derives from $r_* \propto a_5$ (Eq.~\ref{eq:sig}) combined with the positive correlation of $a_5$ with $|\Phi|$; in the case with screening, however, the signal vanishes for $|\Phi|/c^2 > |\Phi_\text{c}|/c^2 = 3.7 \times 10^{-6}$.

Many chameleon constraints have focused on $f(R)$ gravity where $\Delta G/G=1/3$; in this case we require $\lambda_C \lesssim 0.5$ Mpc ($1\sigma$), or equivalently $f_{R0} \equiv df/dR|_{R_0} < \: \text{few} \times 10^{-8}$, where $R_0$ is the current cosmological value of the Ricci scalar. This is stronger than cluster and cosmology constraints by two orders of magnitude~\citep{Song, Schmidt, Yamamoto, Ferraro, Lombriser_2, Lombriser_1, Lombriser_rev, Terukina, Dossett, Wilcox} and by distance indicators~\citep{Sakstein} and rotation curves~\citep{Vikram_RC} by one, and operates in a fully complementary regime to laboratory fifth-force searches~\citep{Adelberger:2003zx, Burrage_rev, Burrage_rev2, Brax}. For $\lambda_C \rightarrow \infty$, which holds for a light scalar field, we expect a $\Delta G/G$ constraint better than $10^{-4}$. These results extend direct constraints on fifth forces from Solar System to galactic scales, helping to fill the gap in the parameter space of tests of gravity~\citep{Baker}. The strength of our bounds owes to the large sample size, great range of gravitational environments probed (including with very low $|\Phi|$), and a vector rather than scalar observable, which effectively affords two orthogonal signals in the plane of the sky.

We have checked that our analysis is converged with number of Monte Carlo realisations, that the AM galaxy--halo connection and smooth density field from BORG are thoroughly sampled, that our MCMC is converged with the number of steps, and that our principal results are insensitive to reasonable variations in $M(<r_*)$ and the assumed uncertainties in galaxy and halo properties.

\section{Caveats and systematic uncertainties}
\label{sec:caveats}

We have marginalised over the statistical uncertainties in most of the model inputs, including the galaxy--halo connection, the smooth density field and the observed properties of galaxies. Nevertheless, we make three key assumptions that may lead to systematic error in our results:

1) We assume that H\textsc{i}-optical offsets generated by non-fifth-force effects follow the Gaussian likelihood model we created for the noise. While baryonic processes such as hydrodynamical drag, ram pressure and stellar feedback may induce a stronger signal than fifth forces, their environment-dependence is unlikely to mimic the effect of screening: our constraints derive primarily from the correlation between the direction of the H\textsc{i}-OC offset and $\vec{a}_5$, as well as both the relative magnitude of these vectors over all galaxies and the precise dependence of the prediction on gravitational potential. Indeed, our model for the uncertainty $\theta$ in the H\textsc{i} centroid implies that on average the entire signal can be accounted for by non-fifth-force effects; that strong constraints are nonetheless attainable attests to the specificity of the features of the signal that fifth forces should induce.

2) To calculate $\Phi$ and $\vec{a}_5$ we assume $\Lambda$CDM structure formation. Although the fifth-force scenarios we investigate would alter cosmology, this is a small effect for $\{\lambda_C, \Delta G\}$ as low as is in question here~\citep{Lombriser_rev}; this systematic error is almost certainly subdominant to the statistical errors in the $\Lambda$CDM galaxy--halo connection and smooth density field. Our method should not therefore be considered a means of probing modified gravity in cosmology, but rather of unearthing any galaxy-scale fifth force in the low-$z$ Universe, of gravitational or non-gravitational origin.

3) Our fiducial noise model sets the positional uncertainty of the H\textsc{i} centroid to be twice as large on a galaxy-by-galaxy basis as the H\textsc{i}-OC displacement itself. If we remove the factor of two in our $\theta$ assignment -- as would roughly be derived by fitting $\theta$ to the data as a zeroth, first or second order polynomial in SNR -- we find $6.6\sigma$ evidence for $\Delta G/G > 0$. This reflects a positive correlation between $\vec{a}_5$ and the observed $\vec{r}_*$ over the unscreened part of the sample across the lower portion of our $\lambda_C$ range ($\lambda_C \lesssim 5$ Mpc), with a maximum log-likelihood at $\lambda_C \simeq 1.8$ Mpc and $\Delta G/G \simeq 0.025$ that is $16$ larger than that obtained by $\Delta G=0$. We describe and validate this possible detection fully in~\citet{Desmond_PRD}, and note that a similar signal is found in~\citet{Desmond_warp}.

\section{Conclusions and outlook}
\label{sec:conc}

We use the observed displacements between galaxies' stellar and gas mass centroids in the \textit{Alfalfa} catalogue to constrain fifth forces that couple differentially to stars, gas and dark matter. As a case study we consider chameleon and symmetron screening, in which stars in otherwise unscreened galaxies self-screen. We deploy the gravitational maps of~\citet{Desmond:2017ctk} to determine screened and unscreened regions of the $d<200$ Mpc Universe, and calculate the acceleration that would be induced at the position of each \textit{Alfalfa} galaxy by a fifth force with strength $\Delta G$ and range $\lambda_C$. Comparing to the data with a Monte Carlo likelihood formalism, we require $\Delta G/G \lesssim 0.1$ for $\lambda_C = 500$ kpc and $\Delta G/G \lesssim \: \text{few} \: \times 10^{-4}$ for $\lambda_C = 50$ Mpc. In $f(R)$ gravity this is $f_{R0} \lesssim \: \text{few} \times 10^{-8}$. The corresponding bounds without screening are $\Delta G/G \lesssim \: \text{few} \: \times 10^{-4}$ and $\Delta G/G \lesssim \: \text{few} \: \times 10^{-3}$. These are the strongest and among the only fifth-force constraints at astrophysical scales.

While our results reveal the gravitational information that can currently be extracted with this signal, they may be strengthened as data from future galaxy surveys is brought to bear. The principal factors limiting the inference in Fig.~\ref{fig:constraint} are the large uncertainty $\theta$ that we use for the angular position of the H\textsc{i} centroid (with average $\bar{\theta} = 36\arcsec$), and the number of galaxies in the sample. To forecast the improvement afforded by future surveys, we generate mock datasets with $N_\text{gal} = f \times N_\text{Alf}$ galaxies ($10^{-3} < f < 1$), and H\textsc{i} angular uncertainty $\Theta \times \theta_i$ ($10^{-3} < \Theta < 1$) for galaxy $i$. We generate a mock signal for each galaxy by randomly scattering around 0 by this uncertainty, and select the galaxies randomly from the full \textit{Alfalfa} sample. We rederive posteriors on $\Delta G/G$ (at $\lambda_C = 5$ Mpc) for each mock dataset, and fit to this data a power-law of the form
\begin{eqnarray}
\sigma\left(\frac{\Delta G}{G}\right)\simeq 8.6 \times 10^{-4} \left(\frac{10^3}{N_\text{gal}}\right)^{0.91}\left(\frac{\bar{\theta}}{1 \ \rm{arcsec}}\right)^{1.00},
\end{eqnarray}
where the left hand side is the $1\sigma$ constraint on $\Delta G/G$. To project constraints for $N_\text{gal} > N_\text{Alf}$ we extrapolate this relation: for $N_\text{gal} \sim 10^8$, $\bar{\theta} \sim 0.1\arcsec$ -- achievable by next-generation radio surveys such as SKA~\citep{SKA0, SKA} -- the constraints on $\Delta G/G$ should be $\mathcal{O}(10^{-9})$. This would be competitive with proposed Solar System tests involving laser ranging to Phobos and optical networks around the Sun~\citep{Sakstein_space}. We caution however that further modelling will be required to extend the gravitational maps to the higher redshift ($z \sim 0.5$) that this $N_\text{gal}$ requires, and also that the time-dependence of parameters such as $f_{R0}$ may impact the inference.

Our analysis is the first to employ ``big data'' from galaxy surveys to constrain gravitational physics with an intra-galaxy signal. We have shown this to afford tighter constraints on fifth forces than other methods involving either cosmological information or cherry-picked astrophysical objects. Nevertheless, the power of tests of this type remains largely unexplored: many more galactic signals -- including disk warps, mass discrepancies, dynamical asymmetries and offsets between kinematics at different wavelengths -- will bring further and independent constraining power. Our work paves the way for fundamental physics to be incorporated as a key science driver in upcoming survey programmes.

\section*{Acknowledgements}

HD is supported by St John's College, Oxford. PGF acknowledges support from Leverhulme, STFC, BIPAC and the ERC.  GL acknowledges support by the ANR grant number ANR-16-CE23-0002 and from the Labex ILP (reference ANR-10-LABX-63) part of the Idex SUPER (ANR-11-IDEX-0004-02). We thank Martha Haynes for sharing the complete \textit{Alfalfa} catalogue before public release, Phil Bull for information on SKA and Jeremy Sakstein and Tessa Baker for comments on the draft. Computations were performed at Oxford and SLAC.

\bibliographystyle{mnras}
\bibliography{GFF}

\bsp

\end{document}